\documentclass[preprint,aps,pre,superscriptaddress]{revtex4-1}
\usepackage[utf8]{inputenc}
\usepackage{natbib}
\usepackage{subfigure}
\usepackage{latexsym}
\usepackage{amsmath}
\usepackage{fullpage}
\usepackage{graphicx}
\usepackage{color}
\usepackage{ulem}
\usepackage{enumerate}

\newcommand{\avg}[1]{\left\langle #1\right\rangle}

\newcommand{\T}[1]{\textrm{#1}}

\begin{document}

\title{Effect of correlations on network controllability}

\author{Márton Pósfai}
\affiliation{Center for Complex Network Research and Department of Physics, Northeastern University, Boston}
\affiliation{Department of Physics of Complex Systems, Eötvös University, Budapest}
\affiliation{Department of Theoretical Physics, Budapest University of Technology and Economics, Budapest}
\affiliation{Center for Network Science, Central European University, Budapest}

\author{Yang-Yu Liu}
\affiliation{Center for Complex Network Research and Department of Physics, Northeastern University, Boston}
\affiliation{Center for Cancer Systems Biology, Dana-Farber Cancer Institute, Boston}

\author{Jean-Jacques Slotine}
\affiliation{Non-linear Systems Laboratory, Department of Mechanical Engineering and Department of Brain and Cognitive Sciences, Massachusetts Institute of Technology, Cambridge}

\author{Albert-László Barabási}
\email{alb@neu.edu}
\affiliation{Center for Complex Network Research and Department of Physics, Northeastern University, Boston}
\affiliation{Department of Medicine, Brigham and Women’s Hospital, Harvard Medical School, Boston}
\affiliation{Center for Network Science, Central European University, Budapest}

\date{\today}

\begin{abstract}
A dynamical system is controllable if by imposing appropriate external signals on a subset of its nodes, it can be driven from any initial state to any desired state in finite time. Here we study the impact of various network characteristics on the minimal number of driver nodes required to control a network. We find that clustering and modularity have no discernible impact, but the symmetries of the underlying matching problem can produce linear, quadratic or no dependence on degree correlation coefficients, depending on the nature of the underlying correlations. The results are supported by numerical simulations and help narrow the observed gap between the predicted and the observed number of driver nodes in real networks.
\end{abstract}

\maketitle

While during the past decade significant efforts have been devoted to understanding the structure, evolution and dynamics of complex networks \cite{Toroczkai-Book-04, Newman-Book-06, Caldarelli-Book-07, Vespignani-Book-09, Cohen-Book-10, Chen-Book-12}, only recently has attention turned to an equally important problem: our ability to control them. Given the problem's importance, recent work has extended the concept of pinning control \cite{Wang-PhysicaA-02, Sorrentino-PhyRevE-07, Gutierrez-SciRep-2012} and structural controllability \cite{nature, Wang-PRE-12, Gang-PRL-12, Vicsek-NaturePhysics-12} to complex networks. Here we focus on the latter approach. A networked system is considered controllable if by imposing appropriate external signals on a subset of its components, called driver nodes, the system can be driven from any initial state to any final state in finite time \cite{Kalman-JSIAM-63, Chui-Book-89, Luenberger-Book-79, Slotine-Book-91}. As the control of a system requires a quantitative description of the governing dynamical rules, progress in this area was limited to small engineered systems. Yet, recently Liu et al. \cite{nature} showed that the identification of the minimal number of driver nodes required to control a network, $N_\T D$, can be derived from the network topology by mapping controllability \cite{Kalman-JSIAM-63} to the maximum matching in directed networks \cite{Lovasz-Book-09}. The mapping indicated that $N_\T{D}$ is mainly determined by the degree distribution $P(k_\T{in},k_\T{out})$. We know, however, that a series of characteristics, from degree correlations \cite{Pastor-Satorras-prl-2001, Maslov-science-2002, Newman-pre-2003} to local clustering \cite{Watts-Nature-98} and communities \cite{Ravasz-Science-02, Palla-Nature-05, Newman-prl-2008, Fortunato-PhysicsReports-10}, cannot be accounted for by $P(k_\T{in},k_\T{out})$ alone, prompting us to ask: which network characteristics affect the system's controllability?

The three most commonly studied deviations from the random network configuration are (i) clustering, manifested as a higher clustering coefficient $C$ than expected based on the degree distribution \cite{BarratWeigt-epjb-2000}; (ii) community structure, representing the agglomeration of nodes into distinct communities, captured by the modularity parameter $Q$ \cite{Newman-prl-2008}; (iii) degree correlations \cite{Foster-pnas-2010}. In Sec.~\ref{sec:motivation} we motivate our work by showing that network characteristics other than the degree distribution also affect network control. In Sec.~\ref{sec:numeric} we use numerical simulations to identify the network characteristics that affect controllability, finding that only degree correlations have a discernible effect. In Sec.~\ref{sec:analytic} we analytically derive $n_\T{D}=N_\T{D}/N$ for random networks with a given degree distribution and correlation profile. More detailed calculations are provided in the Supplementary Information Sec.~III. In Sec.~\ref{sec:real} we test our predictions on real networks. Finally, Sec.~\ref{sec:sum} summarizes our results.

\section{Results}

\subsection{Prediction based on the degree distribution}\label{sec:motivation}

\begin{figure}[h]
	\centering
	\includegraphics[scale=.50]{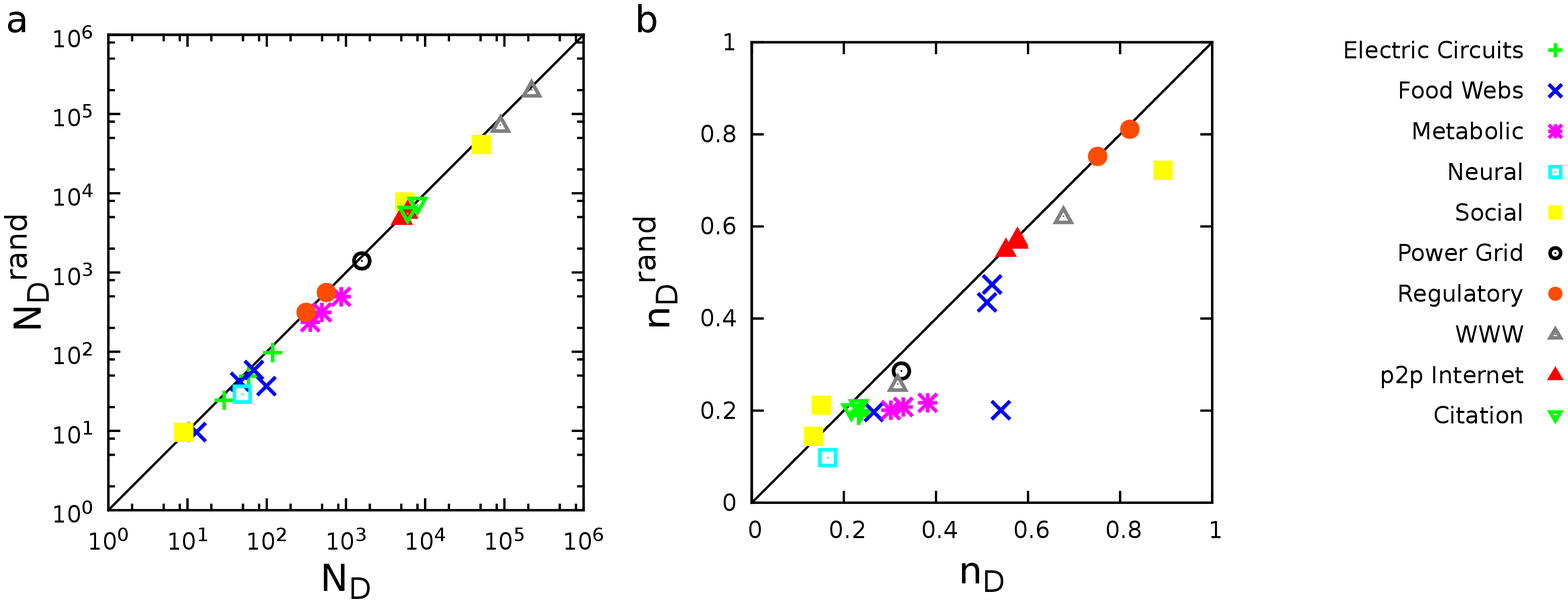}
	\caption{\label{fig:real-rand}(a) We compare $N_\T D$ for real systems to $N_\T D^{\T{rand}}$, representing the number of driver nodes needed to control their randomized counterparts. Randomization eliminates all local and global correlations, only preserving the degree sequence of the original system. We find that the degree sequence predicts the order of magnitude of $N_\T{D}$ correctly, however, small deviations are hidden by the log scale, needed to show the whole span of $N_\T D$ seen in real systems. (b) These deviations are more obvious if we compare the density of driver nodes $n_\T D = N_\T D/N$ and $n_\T D^{\T{rand}}$ in linear scale, finding that for some systems (e.g. regulatory and p2p Internet networks) the degree sequence serves as a good predictor of $n_\T D$, while for other systems (e.g. metabolic networks and food webs) $n_\T D$ deviates from the prediction based solely on the degree sequence.}
\end{figure}

To motivate our study we compared the observed $N_\T D$ to the prediction based on the degree sequence for several real networks. For this we randomize each network preserving its degree sequence and we calculate $N_{\T{D}}^{\T{rand}}$, the number of driver nodes for the randomized network. Plotting $N_\T D$ versus $N_{\T{D}}^{\T{rand}}$ on log-log scale indicates that the degree sequence correctly predicts the order of magnitude of $N_\T{D}$ despite known correlations \cite{Pastor-Satorras-prl-2001, Maslov-science-2002} (Fig.~\ref{fig:real-rand}a). However, by plotting $n_\T D = N_\T D / N$ versus $n_\T D^{\T{rand}}=N_\T D^{\T{rand}}/N$ we observe clear deviations from the degree based prediction (Fig.~\ref{fig:real-rand}b). Our goal is to understand the origin of these deviations, and the degree to which network correlations can explain the observed $n_\T D$.

\subsection{Numerical simulations}\label{sec:numeric}

We start from a directed network with Poisson \cite{Erdos-PMIHAS-60, Bollobas-Book-01} or scale-free degree distribution \cite{Barabasi-Science-99, Albert-RevModPhys-02}. The scale-free network is generated by the static model described in Sec.~\ref{sec:scale-free_model}. We use simulated annealing to add various network characteristics by link rewiring, while leaving the in- and out-degrees unchanged, tuning each measure to a desired value, for details see Sec.~\ref{sec:rewire}. We computed $n_\T{D}$ using the Hopcroft-Karp algorithm \cite{HopcroftKarp-SIAM-1973}.

\begin{figure}[h]
	\centering
	\includegraphics[scale=.35]{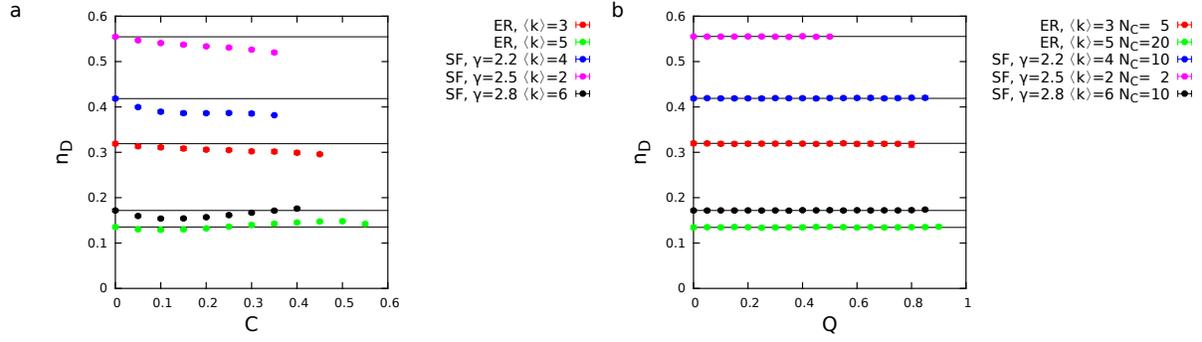}
	\caption{\label{fig:eff_clust_and_mod} Effect of the clustering coefficient $C$ and modularity $Q$ on the density of driver nodes, $n_\T D$. Network size is $N=10,000$. Each data point is an average over 50 independent runs; the error bars, typically smaller than the symbol size, represent the standard deviation of the measurements.}
\end{figure}

\paragraph*{Clustering.}
We use the global clustering coefficient \cite{BarratWeigt-epjb-2000} defined for directed networks as
\begin{align}
 C = \frac{3\cdot \textrm{ number of triangles}}{2\cdot\textrm{number of adjacent edge pairs}}.
\end{align}
The simulations indicate that changes in $C$ only slightly alter $n_\T D$ and that the effect is not systematic (Fig.~\ref{fig:eff_clust_and_mod}a). Hence we conclude that $C$ plays a negligible role in determining $n_\T D$.

\paragraph*{Modularity.}
We quantify the community structure using \cite{Newman-prl-2008, Fortunato-PhysicsReports-10}:
\begin{equation}
 Q = \frac{1}{E}\sum\limits_{vw}\left[ A_{vw} - \frac{k^{(\T{in})}_vk^{(\T{out})}_w}{E}\right]\delta_{c_v,c_w},
\end{equation}
where $A_{vw}$ is the adjacency matrix, $c_v$ and $c_w$ are the communities the $v$ and $w$ nodes belong to, respectively. Specifying $Q$ still leaves a great amount of freedom in the number and size of the communities. We therefore choose to randomly divide the nodes into $N_\T C$ equally sized groups, and increase the edge density within these groups, elevating $Q$ to the desired value.

The simulations indicate that this community structure has no effect on $n_\T D$ (Fig.~\ref{fig:eff_clust_and_mod}b). While adding communities to networks can be achieved in many different ways, and the effect of modularity can be explored in more detail (e.g. hierarchical organization of communities \cite{Ravasz-Science-02, Ravasz-pre-03, Vicsek-plosone-12}, overlapping community structure \cite{Palla-Nature-05, Ahn-Nature-10}, etc), we have failed to detect systematic, modularity induced changes in $n_\T D$, prompting us to conclude that $Q$ does not play a leading role in $n_\T D$.

\paragraph*{Degree correlations.}
\label{sec:sim-degcorr}

\begin{figure}[h]
	\centering
	\includegraphics[scale=.50]{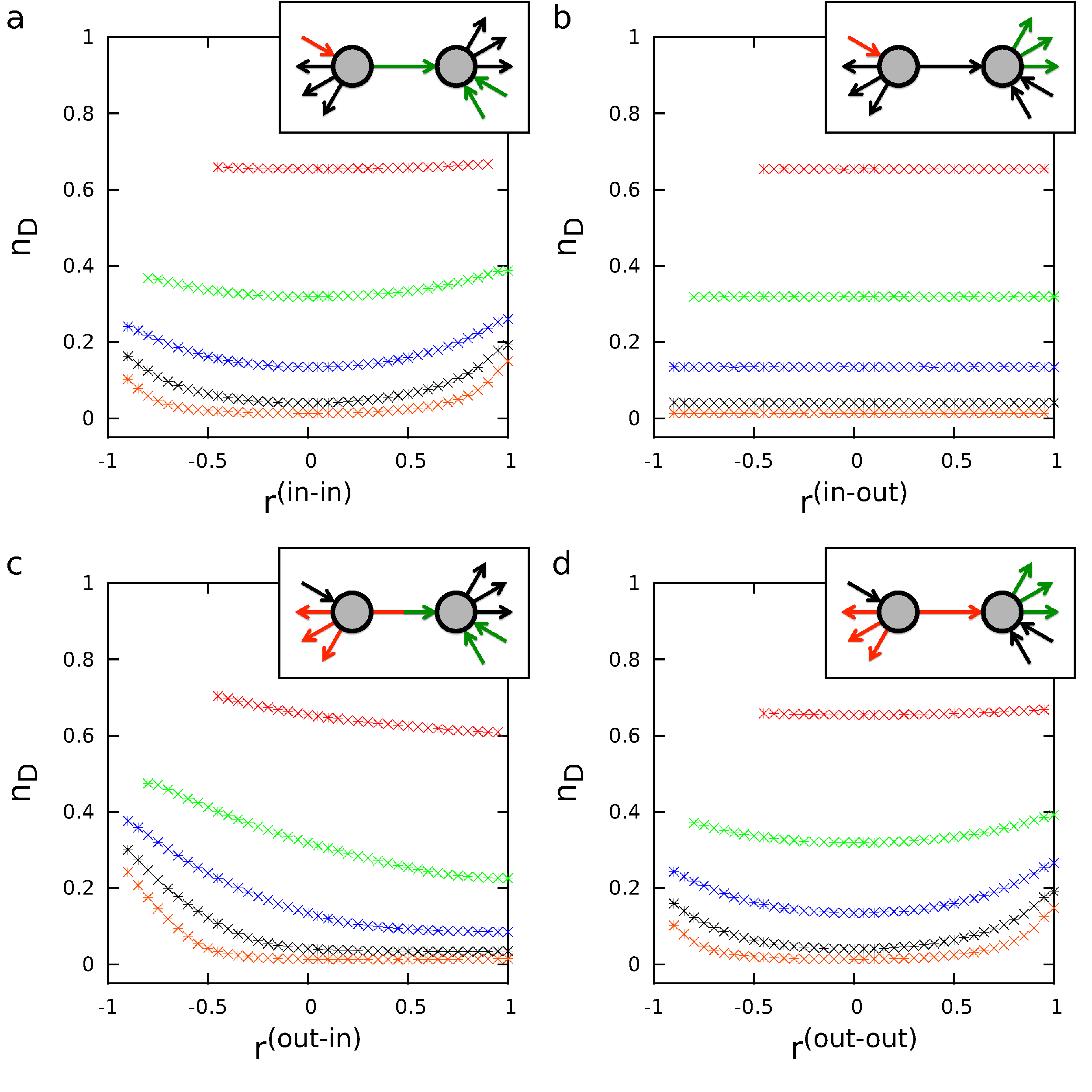}
	\caption{\label{fig:ER_simulation} The impact of degree-degree correlations on the density of driver nodes ($n_\T D$) for the Erdős-Rényi model ($N = 10,000$) for average degrees $\langle k \rangle =1$ (red), $\langle k \rangle =3$ (green),$\langle k \rangle =5$ (blue), $\langle k \rangle =7$ (black) and $\langle k \rangle =9$ (orange). The results are similar for the scale-free model (see Fig.~\ref{fig:SF_simulation}). Each data point is an average of 100 independent runs.}
\end{figure}

\begin{figure}[h]
	\centering
	\includegraphics[scale=.5]{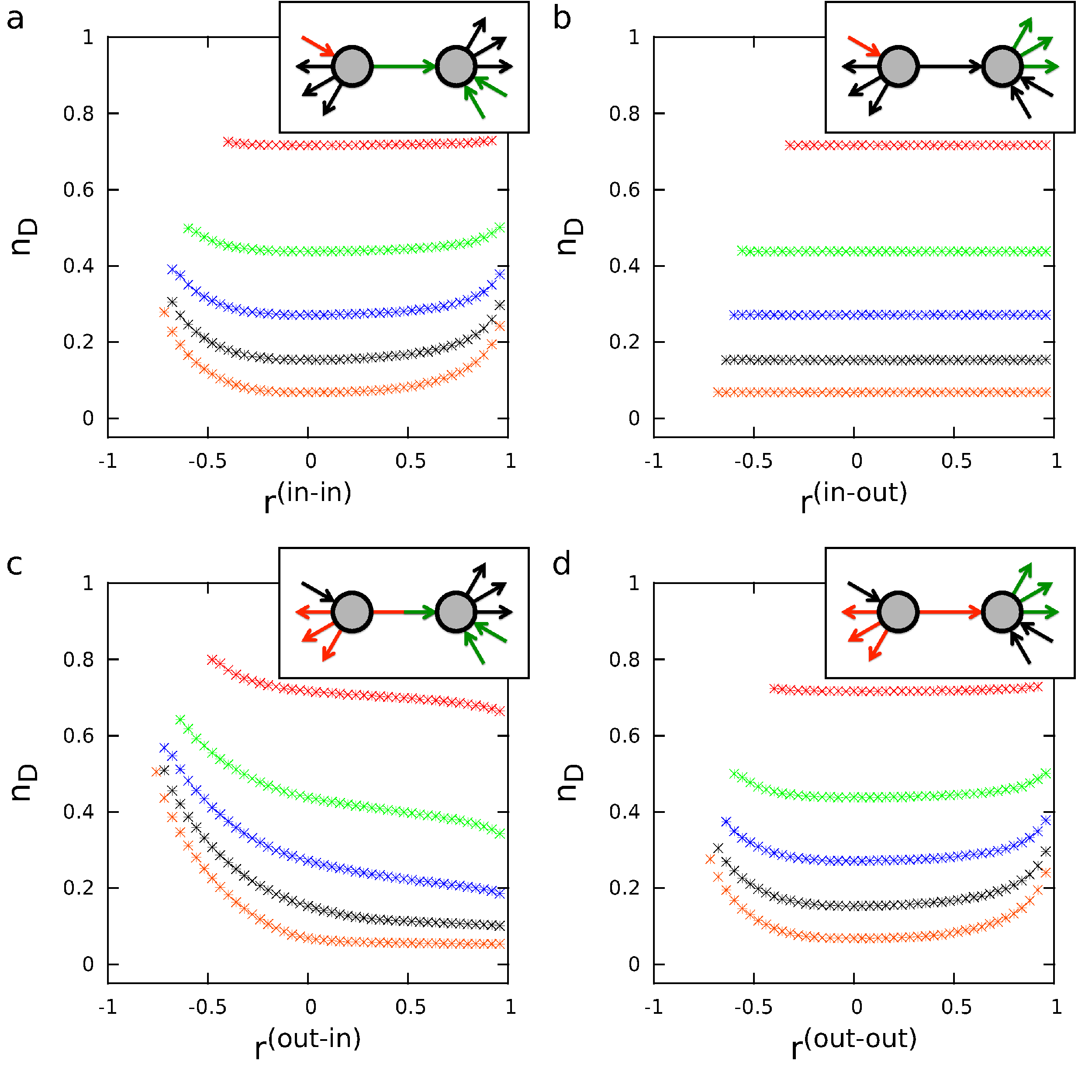}
	\caption{\label{fig:SF_simulation} The impact of degree-degree correlations on the density of driver nodes ($n_\T D$) for the scale-free model ($N = 10,000$, $\gamma=2.5$) for average degrees $\langle k \rangle =1$ (red), $\langle k \rangle =3$ (green),$\langle k \rangle =5$ (blue), $\langle k \rangle =7$ (black) and $\langle k \rangle =9$ (orange). The results are similar for the Erdős-Rényi model (see Fig.~\ref{fig:ER_simulation}). Each data point is an average of 100 independent runs.}
\end{figure}

In directed networks each node has an in-degree ($k_\T i$) and an out-degree ($k_\T o$), thus we can define four correlation coefficients: correlations between the source node's in- and out-degree, and the target node's in- and out-degree (Figs.~\ref{fig:ER_simulation}, \ref{fig:SF_simulation}) \cite{Foster-pnas-2010}. We use the Pearson coefficient to quantify each correlation with a single parameter: 
\begin{align}\label{eq:r_definition}
 r^{(\alpha-\beta)} = \frac{\frac{1}{E} \sum_e \left(k^{(\alpha)}_e-\overline{k^{(\alpha)}}\right) \left(j^{(\beta)}_e-\overline{j^{(\beta)}}\right)}{\sigma^{(\alpha)} \sigma^{(\beta)}},
\end{align}
where $\sum_e\cdot$ sums over all edges, $\alpha,\beta \in \left\{ \T{in, out}\right\}$ is the degree type, $k^{(\alpha)}$ is the degree of the source node, $j^{(\beta)}$ is the degree of the target node. And $\overline{j_\alpha}=\frac{1}{E}\sum_e j^{\alpha}_e$ is the average degree of the nodes at the beginning of each  link, $\sigma_\alpha^2= \frac{1}{E}\sum_e \left(k^{(\alpha)}_e-\overline{k^{(\alpha)}}\right)^2$ is the variance; $\overline{k^{(\beta)}}$ and $\sigma^{(\beta)}$ are defined similarly.

Simulations shown in Figs.~\ref{fig:ER_simulation} and \ref{fig:SF_simulation} indicate that degree correlations systematically affect $n_\T D$. We observe three distinct types of behavior:
\begin{enumerate}[(i)]
 \item $n_\T D$ depends monotonically on $r^{(\T{out-in})}$, so that low (negative) correlations increase $n_\T D$ and high (positive) correlations lower $n_\T D$ (Figs.~\ref{fig:ER_simulation}c, \ref{fig:SF_simulation}c); 
 \item Both $r^{(\T{in-in})}$ and $r^{(\T{out-out})}$ increase $n_\T D$, independent of the sign of the correlations (Figs.~\ref{fig:ER_simulation}a, \ref{fig:ER_simulation}d, \ref{fig:SF_simulation}a, \ref{fig:SF_simulation}d);
 \item $r^{(\T{in-out})}$ has no effect on $n_\T D$ (Figs.~\ref{fig:ER_simulation}c, \ref{fig:SF_simulation}c).
\end{enumerate}
The behavior is qualitatively the same for Erdős-Rényi (Fig.~\ref{fig:ER_simulation}) and scale-free (Fig.~\ref{fig:SF_simulation}) networks.

The diversity of these numerical results require a deeper explanation. Therefore in the remaining of the paper we focus on understanding analytically the role of degree correlations, which, by systematically altering $n_\T D$, affect the system's controllability.

\subsection{Analytical framework}\label{sec:analytic}

The task of identifying the driver nodes can be mapped to the problem of finding a maximum matching of the network \cite{nature}. A matching is a subset of links that do not share start or end points. We call a node matched if a link in the matching points at it and we gain full control over a network  if we control the unmatched nodes. The cavity method has been successfully used to calculate the size of the maximum matching for undirected \cite{Mezard-jsm-2006} and directed \cite{ nature} network ensembles with given degree distribution. Here we study network ensembles with a given degree correlation profile.

We calculate $n_\T D$ analytically for a given $P(k_{\T{in}}, k_{\T{out}})$ and selected degree-degree correlation $e(j_{\T{in}},j_{\T{out}};k_{\T{in}},k_{\T{out}})$, representing the probability of a directed link pointing from a node with degrees $j_{\T{in}}$ and $j_{\T{out}}$ to a node with degrees $k_{\T{in}}$ and $k_{\T{out}}$. In the absence of degree correlations (neutral case)
\begin{equation}
e^{(0)}(j_\T{i},j_\T{o};k_\T{i},k_\T{o})=P^{\T{(in)}}(j_\T{i})Q^{\T{(out)}}(j_\T{o}) Q^{\T{(in)}}(k_\T{i})P^{\T{(out)}}(k_\T{o}),
\end{equation}
where $Q^{\T{(out)}}(j_\T{o})=\frac{2j_\T{o}}{\avg{k}}P^{\T{(out)}}(j_\T{o})$, $Q^{\T{(in)}}(k_\T{i})=\frac{2k_\T{i}}{\avg{k}}P^{\T{(in)}}(k_\T{i})$ and $\avg{k}$ is the average degree. To ensure analytical tractability we chose \cite{Newman-pre-2003}
\begin{subequations}\label{eq:eperturb}
\begin{align}
 e^{(\T{in-in})}(j_\T{i},j_\T{o};k_\T{i},k_\T{o})&=Q^{\T{(out)}}(j_\T{o})P^{\T{(out)}}(k_\T{o})\left[P^{\T{(in)}}(j_\T{i})Q^{\T{(in)}}(k_\T{i}) + r^{(\T{in-in})} m^{(\T{in-in})}(j_\T{i},k_\T{i}) \right], \label{eq:ii}\\
 e^{(\T{in-out})}(j_\T{i},j_\T{o};k_\T{i},k_\T{o}) &= Q^{\T{(out)}}(j_\T{o})Q^{\T{(in)}}(k_\T{i}) \left[P^{\T{(in)}}(j_\T{i})P^{\T{(out)}}(k_\T{o}) + r^{(\T{in-out})} m^{(\T{in-out})}(j_\T{i},k_\T{o}) \right],\label{eq:io}\\
 e^{(\T{out-in})}(j_\T{i},j_\T{o};k_\T{i},k_\T{o}) &=P^{\T{(in)}}(j_\T{i})P^{\T{(out)}}(k_\T{o})\left[Q^{\T{(out)}}(j_\T{o})Q^{\T{(in)}}(k_\T{i}) + r^{(\T{out-in})} m^{(\T{out-in})}(j_\T{o},k_\T{i}) \right],\label{eq:oi}\\
 e^{(\T{out-out})}(j_\T i,j_\T o;k_\T i,k_\T o)&=Q^{\T{(in)}}(j_\T i)P^{\T{(in)}}(k_\T i)\left[ P^{\T{(out)}}(j_\T o)Q^{\T{(out)}}(k_\T o) + r^{(\T{out-out})}  m^{(\T{out-out})}(j_\T o,k_\T o) \right].\label{eq:oo}
\end{align}
\end{subequations}
 By fixing $m^{(\alpha-\beta)}(j,k)$ ($\alpha,\beta \in \{\T{in},\T{out}\}$) we obtain a one parameter network ensemble characterized by $r^{(\alpha-\beta)}$, where $m^{(\alpha-\beta)}(j,k)$ satisfies the constraints
\begin{align}
 \sum_{j=0}^{\infty} m^{(\alpha-\beta)}(j,k) = \sum_{k=0}^\infty m^{(\alpha-\beta)}(j,k) = 0, \\
 \sigma^{(\alpha)}\sigma^{(\beta)} \sum_{j,k=0}^{\infty} jk\cdot m^{(\alpha-\beta)}(j,k) = 1,
\end{align}
and all elements of $e^{(\alpha-\beta)}(j,k)$ are between 0 and 1.

Our goal is to understand the relation between $n_\T D$ and the degree correlation coefficient $r^{(\alpha-\beta)}$. Assuming that $r^{(\alpha-\beta)}$ is small we treat the correlations as perturbations to the neutral case, discussing the impact of the four $r^{(\alpha-\beta)}$ correlations separately.

{\it Out-in correlations:} Using equation~(\ref{eq:oi}) and keeping the first nonzero correction we obtain (Supplementary Information Sec.~III.):
\begin{equation}\label{eq:final_oi}
 \overline{n_\T{D}}^{(\T{out-in})} = \overline{n_\T{D}}^{(0)} - r^{(\T{out-in})} \frac{\avg{k}}{4}\left[ M_1(\hat w_2,1-w_1) + M_1(1-\hat w_1,w_2) \right],
\end{equation}
where $\overline{n_\T{D}}^{(0)}$ is the fraction of driver nodes of the uncorrelated network; $w_i$ and $\hat w_i$ only depend on $P(k_{\T{in}}, k_{\T{out}})$ \cite{nature}, and
\begin{equation}\label{eq:M1}
 M_1(x,y) = \sum_{j,k=1}^{\infty} m^{\T{(out-in)}}(j,k)x^{j-1}y^{k-1}.
\end{equation}

Equation~(\ref{eq:final_oi}) predicts that $\overline{n_\T{D}}$ depends linearly on $r^{(\T{out-in})}$, a prediction supported by simulations for small $r^{(\T{out-in})}$ (Figs.~\ref{fig:ER_simulation}c and \ref{fig:SF_simulation}c). This behavior is also revealed by the equivalent problem of finding the maximum matching of graphs \cite{nature}. For a node $A$ with out-degree $k_0$, by definition only one edge can be in the matching. If the remainder $k_0-1$ edges point to nodes with degree $1$ (disassortative case), $A$ inhibits them from being matched, so we have to control each of them individually, increasing $n_{\T D}$. If the remainder $k_0-1$ edges point to hubs (assortative case), these hubs are likely to be matched through another incoming edge, decreasing $n_{\T D}$.

{\it Out-out correlations:} The cavity method indicates that for out-out correlations the first nonzero correction is of order $\left( r^{(\T{out-out})}\right)^2$:
\begin{equation}
\label{eq:final_oo}
\overline{n_\T{D}}^{(\T{out-out})} = \overline{n_\T{D}}^{(0)} + {r^{(\T{out-out})}}^2 \frac{\avg{k}}{8}\left[ H^{\T{(in)}\prime}(1-w_1)M_2(\hat w_2) + H^{\T{(out)}\prime}(w_2)M_2(1-\hat w_1) \right],
\end{equation}
where $H^{(\alpha)}(x)=\sum_{k=1}^{\infty} Q^{(\alpha)}(k)x^{k-1}$ ($\alpha \in \left\{ \T{in, out}\right\}$) only depends on $P(k_{\T{in}}, k_{\T{out}})$ and
\begin{align}\label{eq:M2}
M_2(x) = \sum_{j,k=1,l=0}^\infty \frac{m^{\T{(out-out)}}(l,j)m^{\T{(out-out)}}(l,k)}{P^{\T{(out)}}(l)} x^{j-1}x^{k-1}.
\end{align}

\begin{figure}[h]
	\centering
	\includegraphics[width=8cm]{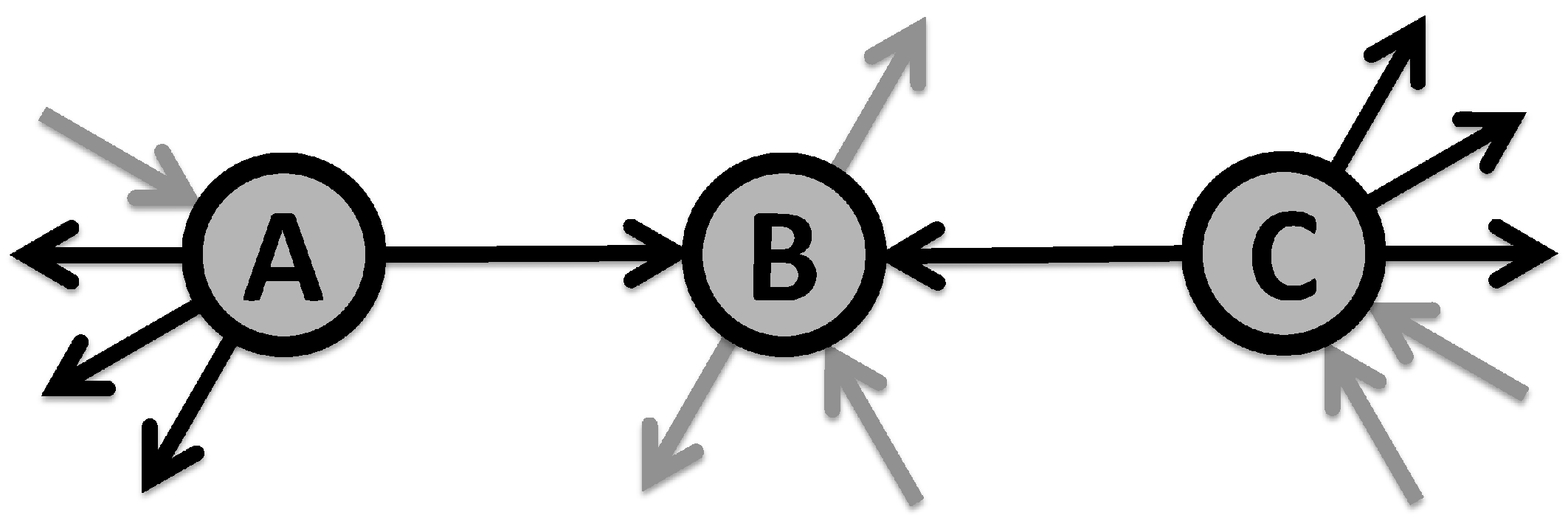}
	\caption{\label{fig:oo_induced_corr} One-step out-out correlations induce positive two-step correlation. Positive (negative) correlation between neighboring nodes means that if node $A$ has high out-degree, then node $B$ is likely to  have high (low) out-degree, and hence $C$ will likely have high out-degree.}
\end{figure}

Equation~(\ref{eq:final_oo}) predicts that $\overline{n_\T{D}}^{\T{(out-out)}}$ does not depend on the out-out correlation of the directly connected nodes, but only on the correlation between the second neighbors, hence its dependence is quadratic in $r^{(\T{out-out})}$, a prediction supported by numerical simulations (Figs.~\ref{fig:ER_simulation}d and \ref{fig:SF_simulation}d). Indeed, positive (negative) $r^{\T{(out-out)}}$ correlation between the immediate neighbors means that if node $A$ has high out-degree, then node $B$ is expected to have high (low) out-degree, and therefore $C$ is likely to have high out-degree (Fig.~\ref{fig:oo_induced_corr}). That is, both positive and negative one-step out-out correlations induce positive two-step correlations, accounting for the symmetry of the effect observed in simulations (Figs.~\ref{fig:ER_simulation}d and \ref{fig:SF_simulation}d).

{\it In-in correlations:} Switching the direction of each link does not change the matching, but turns out-out correlations into in-in correlations. So $\overline{n_\T{D}}^{\T{in-in}}$ can be obtained by exchanging $P^{\T{(in)}}(k_{\T{in}})$ and $P^{\T{(out)}}(k_{\T{out}})$ in equation~(\ref{eq:final_oo}), predicting again a quadratic dependence on $r^{\T{(in-in)}}$, supported by the numerical simulations (Figs.~\ref{fig:ER_simulation}a and \ref{fig:SF_simulation}a).

{\it In-out correlations:} The equations for $\overline{n_\T D}$ do not depend on the in-degree of the source and the out-degree of the target of a link, hence we predict that $r^{\T{(in-out)}}$ does not play a role in network controllability, a prediction supported by the simulations (see Figs.~\ref{fig:ER_simulation}b and \ref{fig:SF_simulation}b).

Taken together, we predict that the functional dependence of $\overline{n_\T{D}}$ on degree correlations defines three classes of behaviors, depending on the matching problem's underlying symmetries: $\overline{n_\T{D}}$ has no dependence on $r^{\T{(in-out)}}$, linear dependence on $r^{\T{(out-in)}}$ and quadratic dependence on $r^{\T{(in-in)}}$ and $r^{\T{(out-out)}}$. These predictions are fully supported by numerical simulations (Figs.~\ref{fig:SF_simulation} and~\ref{fig:SF_simulation}): for small $r$ we see no dependence on $r^{\T{(in-out)}}$, an asymmetric, monotonic dependence on $r^{\T{(out-in)}}$, and a symmetric on $r^{\T{(in-in)}}$ and $r^{\T{(out-out)}}$.

\begin{figure}[h]
	\centering
	\includegraphics[scale=.5]{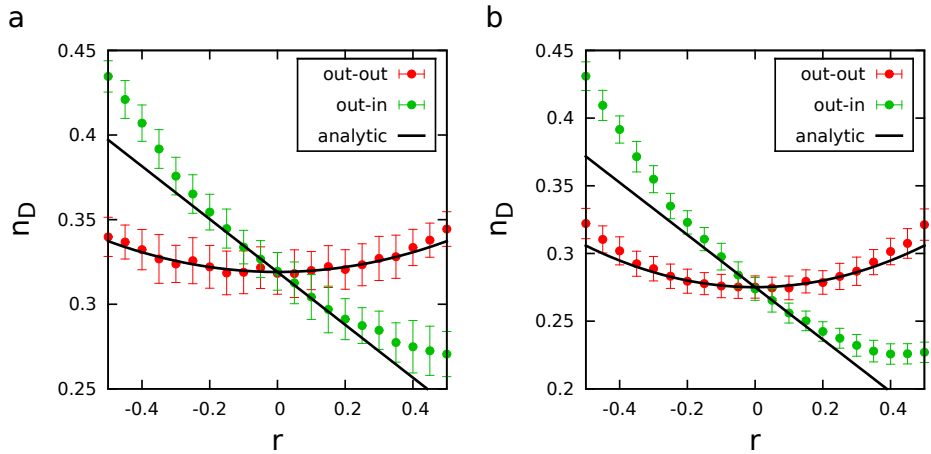}
	\caption{\label{fig:pert_test} The analytic formulas are tested with simulations on an (a) Erdős-Rényi model and on a (b) scale-free model. We used the algorithm proposed in \cite{Newman-pre-2003} to set $e^{(\alpha-\beta)}(j_\T{i},j_\T{o};k_\T{i},k_\T{o})$. For (a) network we choose $N=1,000$ and $\langle k \rangle=3$; for (b) $N=1,000$, $\gamma=2.5$ and $\langle k \rangle=4$. Each data point is an average over 100 independent runs; the errors represent by the standard deviation of the measurements.}
\end{figure}

To directly compare the analytical predictions to simulations we need to know the complete $e(j_\T{i},j_\T{o};k_\T{i},k_\T{o})$ distribution, which is not explicitly set in the simulations in Sec.~\ref{sec:numeric}. So to test the results we use a rewiring method that sets the $e(j_\T{i},j_\T{o};k_\T{i},k_\T{o})$ distribution, not only the $r$ correlation coefficient \cite{Newman-pre-2003}. This method is not as robust as our original algorithm and the range of accessible $r$ values is more restricted. However, since our results are based on perturbation scheme we only expect them to be correct for small $r$ values. Indeed, we find that the predictions quantitatively reproduce the numerical results in a fair interval of $r^{(\alpha-\beta)}$ (Fig.~\ref{fig:pert_test}).

\subsection{Real networks}\label{sec:real}

We test the predictions provided by the developed analytical and numerical tools on a set of publicly available network datasets. When complex systems are mapped to networks, the links connecting the nodes represent interactions between them. In this context self-loops represent self-interactions, with a strong, well understood impact on controllability~\cite{nature, Cowan-arXiv-2011}.  While in some systems self-loops are obviously present (e.g. neural networks), in others they are manifestly absent (e.g. electric circuits~\cite{Lin-IEEE-74}). Our purpose here is to test the effect of correlations, hence we rely on datasets that capture the wiring diagram of various complex systems with different correlation properties. Therefore, even if in a few of these maps self-loops are missing, it is beyond the scope of this work to complete these networks. However, when studying controllability of a particular system, careful thought has to be put into whether self-loops are present or not. We present a systematic study on the effect of self-loops in Supplementary Information Sec.~II.B.

\begin{figure}[h]
	\centering
	\includegraphics[width=.9\textwidth]{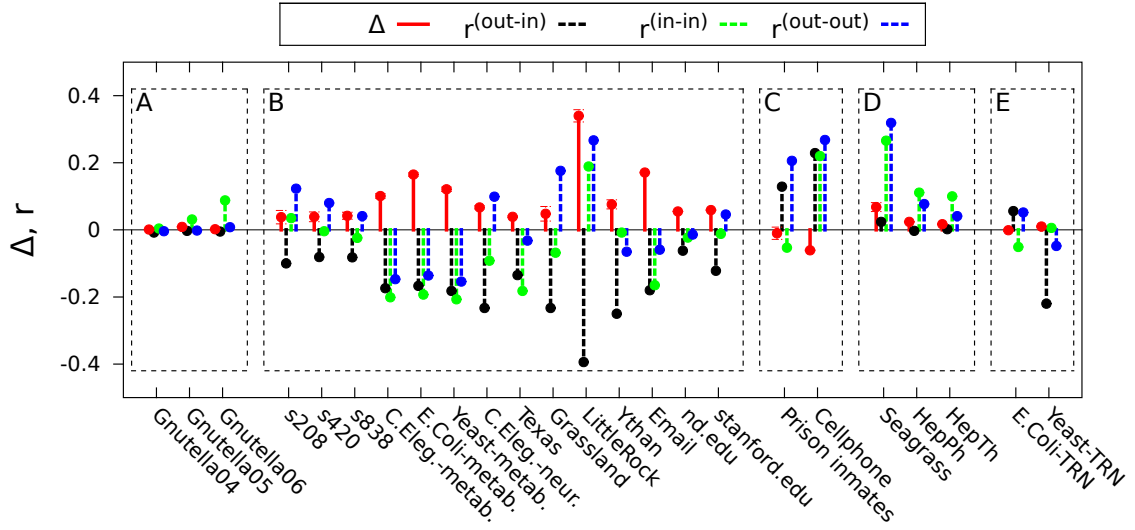}
	\caption{\label{fig:real-summary} The observed and predicted deviation between $N_{\T{D}}$ and $N_{\T{D}}^{\T{rand}}$. Red line: $\Delta= \left(N_{\T{D}} - N_{\T{D}}^{\T{rand}}\right)/N$, the prediction error based on the degree sequence. Dashed lines: correlations relevant to controllability. For each network $\Delta$ is calculated by averaging over 50 independent configurations.}
\end{figure}

To test the impact of our predictions on real networks we calculate
\begin{equation}
\Delta = \frac{N_{\T{D}} - N_{\T{D}}^{\T{rand}}}{N}, 
\end{equation}
where $N_{\T{D}}^{\T{rand}}$ represent the number of driver nodes for the degree-preserved randomized version of the original network. Hence if $\Delta = 0$ then $P(k_\T{in},k_\T{out})$ accurately determines $N_\T{D}$;  if $\Delta \neq 0$ then the structural properties not captured by the degree sequence influence its controllability. We measure the correlations in several real networks and based on our numerical and analytical results we predict the sign of $\Delta$ (Fig.~\ref{fig:real-summary}). We grouped the networks according to our predictions. We provide the details of each network dataset in the Supplementary Information Table~SI.

\paragraph*{Group A.}
The networks of p2p Internet (Gnutella filesharing clients) do not have strong correlations, therefore we expect $n_\T{D}$ to be correctly approximated by the prediction based on $P(k_\T{in}, k_{\T{out}})$ (i.e. $\Delta\approx0$), in line with the empirical observations.

\paragraph*{Group B.}
As in most networks the three relevant correlations coexist to some degree (Fig.~\ref{fig:real-summary}), it is impossible to isolate their individual role. Yet, the networks in this group (electric circuits, metabolic networks, neural networks, power grids and food webs with exception of the Seagrass network) all have negative out-in and nonzero in-in and out-out correlations, each of which individually increase $n_\T{D}$ as we showed above. Therefore we predict $\Delta>0$, in line with the empirical observations.

\paragraph*{Group C.}
Only the prison social-trust and the cell phone network feature significant positive out-in correlations. These networks also display nonzero in-in and out-out correlation, leading to the coexistence of two competing effects: out-in correlations decrease $n_\T D$ and the out-out and in-in correlations increase $n_\T{D}$. Since the out-in correlation is a first order effect (equation~(\ref{eq:final_oi})), while out-out and in-in correlations are only of second order (equation~(\ref{eq:final_oo})), we expect a decrease in $n_\T D$ (i.e. $\Delta<0$), consistent with the empirical results.

\paragraph*{Group D.}
The Seagrass food web and citation networks do not feature significant out-in correlations, only the secondary in-in and out-out correlations, hence we expect $n_\T{D}$ to increase ($\Delta>0$), consistent with the observations.

\paragraph*{Group E.}
Only the transcriptional regulatory networks are somewhat puzzling in that they show degree correlations, yet the degree sequence still correctly gives $n_\T{D}$. However, the simulations indicated that the effect of correlations is negligible for high $n_\T D$. And  our analytical results showed that the value of the correction depends on details of $e(j_\T{i},j_\T{o};k_\T{i},k_\T{o})$, not captured by the Pearson coefficient $r$. These observations highlight that even though in most cases our qualitative predictions based on $r$ are valid, in some cases further investigation is required.

\section{Discussion}\label{sec:sum}

The goal of our paper was to clarify the higher order network characteristics that influence controllability. We studied the effect of three topological characteristics: clustering, modularity and degree correlations. We used numerical simulations to identify the role of the relevant characteristics, finding that changes in the clustering coefficient and the community structure have no systematic effect on the the minimum number of driver nodes $n_\T D$. In contrast degree correlations showed a robust effect, whose magnitude and direction depends on the type of correlation. Using the cavity method we derived $n_\T D$ for networks with given degree distribution and correlation profiles, finding results that are consistent with our numerical simulations. For real networks these numerical and analytic results enabled us to qualitatively explain the deviation of the observed $n_\T D$ from the prediction based only on $P(k_\T{in}, k_\T{out})$.

Our results not only offer a new perspective on the role of topological properties on network controllability, but also raise several questions. Future research directions include determining the optimal network structure to minimize the number of necessary driver nodes, and studying how different network characteristics influence the robustness of the control configuration.

\section{Methods}\label{sec:methods}

\subsection{Generating a scale-free network}
\label{sec:scale-free_model}

We use the static model to generate directed scale-free networks~\cite{Goh-PRL-2001}. We start from $N$ disconnected nodes and assign a weight $w_i = (i+i_0)^{-\alpha}$ to each node $i$ ($i=1\ldots N$). We randomly select two nodes $i$ and $j$ with probability proportional to $w_i$ and $w_j$ respectively and if they are yet not connected, we connect them. We allow self-loops, but avoid multi-edges. We repeat the process until $L$ links have been placed. The resulting network has average degree $\langle k \rangle = 2L/N$, and $P^\T{(in/out)}(k)\sim k^{-\gamma}$ for large $k$, where $\gamma = 1 + \frac{1}{\alpha}$, and maximum degree $k_{\T{max}}\sim i_0^{-\alpha}$.

To systematically study correlations, the starting network has to be uncorrelated. However, the presence of hubs may induce unwanted degree correlations~\cite{Vespignani-epjb-2006}, and may also considerably limit the maximum and minimum correlations accessible via rewiring \cite{Menche-PRE-2010}. We overcome these difficulties by introducing a structural cutoff in the degrees, choosing $i_0$ to ensure $k_{\T{max}} <\left( \langle k \rangle N\right)^{1/2}$~\cite{ChungLu-AnnalsofComb-2002}. Note, that in the static model of Goh et al. $i_0=0$~\cite{Goh-PRL-2001}. 

As both in- and out-degree of node $i$ is proportional to $w_i$, the above procedure results in correlations between the in- and out-degrees of node $i$. To eliminate the correlations, we randomize the in-degree sequence while keeping the out-degree sequence unchanged. 

\subsection{Rewiring algorithm}
\label{sec:rewire}

We use degree preserving rewiring~\cite{Maslov-science-2002} to add each network characteristic. Suppose that the chosen network characteristic is quantified by a metric $X$. To set its value to $X^*$, we define the $E(X)=\lvert X-X^*\rvert$ energy, so $E(X^*)$ is a global minimum. We minimize this energy by simulated annealing~\cite{numrec-book-1992}: (1) choose two links at random with uniform probability; (2) rewire the two links and calculate the energy $E(X)$ of the resulted network; (3) accept the new configuration with probability
\begin{equation}
 p = \begin{cases}
    1, &\text{if } \Delta E \leq 0\\
    e^{-\beta\Delta E}, &\text{if } \Delta E > 0,
  \end{cases},
\end{equation}
where the $\beta$ parameter is the inverse temperature; (4) repeat from step one and gradually increase $\beta$. Stop if $\left\lvert E(X)-E(X^*) \right\rvert$ is smaller than a predefined value.

Note, that keeping the degree sequence bounds the possible values of $X$ that can be reached by rewiring. In all cases we study the full interval of accessible $X$ values.

\section*{Acknowledgements}

This work was supported by the Network Science Collaborative Technology Alliance sponsored by the US Army Research Laboratory under Agreement Number W911NF-09-2-0053; the Defense Advanced Research Projects Agency under Agreement Number 11645021; the Defense Threat Reduction Agency award WMD BRBAA07-J-2-0035; FET IP project MULTIPLEX (3A532) and the generous support of Lockheed Martin. M.~Pósfai has received funding from the European Union Seventh Framework Programme (FP7/2007-2013) under grant agreement No. 270833.

\section*{Contributions}

All authors designed and did the research. M.P. analysed the empirical data and did the analytical and numerical calculations. A.-L.B. was the lead writer.

\section*{Competing financial interests}

The authors declare no competing financial interests.

\pagebreak

\end{document}